\documentclass[pra,twocolumn,superscriptaddress,longbibliography,amsmath,amssymb]{revtex4-1}
 
\usepackage{graphicx}
\usepackage{color}

\usepackage{hyperref}
\hypersetup{
    colorlinks=true,
    linkcolor=red,
    filecolor=magenta,      
    urlcolor=cyan,
    }

\usepackage{dcolumn}
\usepackage{bm}

\begin{document}

\title{Quantum Scattering of Fullerene $^{12}$C$_{60}$ with Rare Gas Atoms and its selection rules for rotational quenching}

\author{Alexander Petrov}
\affiliation{Department of Physics, Temple University, Philadelphia, Pennsylvania 19122, USA}
\author{Anna Linnik}
\affiliation{Department of Physics, Temple University, Philadelphia, Pennsylvania 19122, USA}
\author{Jacek K{\l}os} 
\affiliation{Department of Physics, Temple University, Philadelphia, Pennsylvania 19122, USA}
\affiliation{Joint Quantum Institute, National Institute of Standards and Technology and University of Maryland, Gaithersburg, Maryland 20899, USA }
\author{Eite Tiesinga} 
\affiliation{Joint Quantum Institute, National Institute of Standards and Technology and University of Maryland, Gaithersburg, Maryland 20899, USA }
\author{Svetlana Kotochigova} 
\email{skotoch@temple.edu}
\affiliation{Department of Physics, Temple University, Philadelphia, Pennsylvania 19122, USA}

\begin{abstract}
The discovery of the C$_{60}$ fullerene opened new horizons to design carbon nanostructures with targeted electronic structure as well as transport and optical properties. For example, endohedral $^{12}$C$_{60}$ molecules were proposed as candidates  for functional quantum architectures to store and manipulate encased atomic and molecular qubits. Recent advances in cryogenic buffer-gas cooling and frequency-comb spectroscopy have enabled rovibrational quantum-state-resolved measurements of gas-phase $^{12}$C$_{60}$, revealing rotational fine structure  reflecting its high icosahedral symmetry.
Here, we present a perturbative quantum description of the $^{12}$C$_{60}$ molecule interacting with a buffer gas of $^{40}$Ar atoms at temperatures of order 150~K, including a detailed analysis of their electronic structure, their interaction anisotropies, and the collision-induced rotational quenching of $^{12}$C$_{60}$ in its vibrational and electronic ground state. The role of the icosahedral symmetry  on the collisional dynamics is emphasized leading to a complex dependence on the $^{12}$C$_{60}$ rotational quantum number. 
Finally, we compute the isotropic and anisotropic static and dynamic dipole polarizability of $^{12}$C$_{60}$ in its
absolute ground state in order to evaluate the long-range, van der Waals interaction between $^{12}$C$_{60}$ and $^{40}$Ar.
\end{abstract}

\maketitle

\section{Introduction}\label{sec:intro}

The  discovery of the buckminsterfullerene, C$_{60}$, by Kroto {\em et al.}~\cite{Kroto1985} in 1985 initiated physico-chemical research of this carbon allotrope and its derivatives~\cite{Kratschmer1990, Dresselhaus1996}.  This molecule, with its highly symmetric icosahedral structure, consists of a closed cage of 12 pentagons and 20 hexagons.  Since then, an extended family of fullerenes,  including C$_{70}$, C$_{74}$, C$_{76}$, C$_{78}$, and C$_{80}$, has been synthesized, each preserving the 12-pentagons, while differing in the number and arrangement of hexagons. The unprecedented properties of this family and its derivatives inspired many applications in nanocarbon chemistry, superconductivity, and optoelectronics \cite{Kausar:2024}. 

Fullerene molecules are still regarded as an exotic molecular form of carbon, even though $^{12}$C$_{60}$  with its zero nuclear spin can  easily be formed in the gas phase \cite{Kroto1985} as well as in a crystal structure~\cite{Ibberson1991}. As fullerenes are nearly spherical in shape,  they can encapsulate single atoms, such as lanthanides~\cite{Heath:1985},  small molecules~\cite{Kurotobi:2011, Bacic:2018, Jarvis:2021}, as well as even larger molecules~\cite{Popov:2013,Vyas:2024}.  Recently,   the hydrocarbon CH$_4$ was  placed  inside  $^{12}$C$_{60}$ fullerenes using  ``molecular surgery" as pioneered by Jafari {\em et al.}~\cite{Jafari:2025}. These authors also performed tera\-hertz spectroscopy on this endohedral fullerene.  Another type of  fullerene derivative is created by binding  dopant atoms or molecules to the outside a fullerene~\cite{Xie:2004,Weng:2010,Filippone:2021}. These derivatives are called exohedral fullerenes. Here, alkali-metal-doped fullerenes have attracted attention due to their superconductive properties~\cite{Tsai:1993,Ramirez:2015,Winkelmann:2009}. Finally, complex-shaped carbon molecules, such as  nanoballs and nanotubes, can be created \cite{Tuli:2023}. 

Small and large carbon-containing molecules as well as clusters have been observed by astrophysicists studying the infrared emission  in the intergalactic medium \cite{Herbig1975, Leger1984, Leger1987a,Leger1987b,Herbig2000}. Fullerenes in these media were first observed by Ref.~\cite{Cami2010} and modeled, for example, in Refs.~\cite{Zhang2013,Omont:2016}. Molecules of the fullerene family were proposed as candidates for functional quantum architectures to store and manipulate atomic and molecular qubits in quantum computing devices \cite{Feng:2004,Changala2019}.

In the gas phase, fullerenes and their derivatives are characterized by a large number of rotational, vibrational and electronic degrees of freedom or normal modes. 
Recently, gas-phase $^{12}$C$_{60}$ was cooled to a temperature of 150 K using buffer gas cooling \cite{Changala2019} 
and high-resolution infrared spectroscopy on a vibrational band resolved hundreds of rotational transitions $J\to J'$.
Here, $J$ and $J'$ are the total molecular angular momenta of the initial and final vibrational state, respectively.
In a second investigation \cite{Liu2022}, these authors describe the interaction of $^{12}$C$_{60}$ with atomic and inert molecular  buffer gasses. They combined state-selective optical pumping on the same vibrational band of $^{12}$C$_{60}$ with buffer gas collisions to observe rotational and vibrational energy transfers. 
The role of the icosahedral symmetry of $^{12}$C$_{60}$ could be deduced from complex level repulsions in $J-J'=0$ transitions, the so-called P-branch~\cite{Lee:2023}.  Regimes with and without level repulsion as a function of $J$ were observed, corresponding to ergodic and non-ergodic behavior, respectively. 
In 2025, an even colder, 30 K $^{12}$C$_{60}$  sample was achieved with a helium buffer gas \cite{Chan:2025}. The authors of this experiment also tested the symmetrization postulate for 60 identical bosonic carbon atoms and its connection to spin statistics. A review of these research efforts can be found in Ref.~\cite{Liu_Review:2025}.

\begin{figure}
\includegraphics[scale=0.4,trim=0 0 0 0,clip]{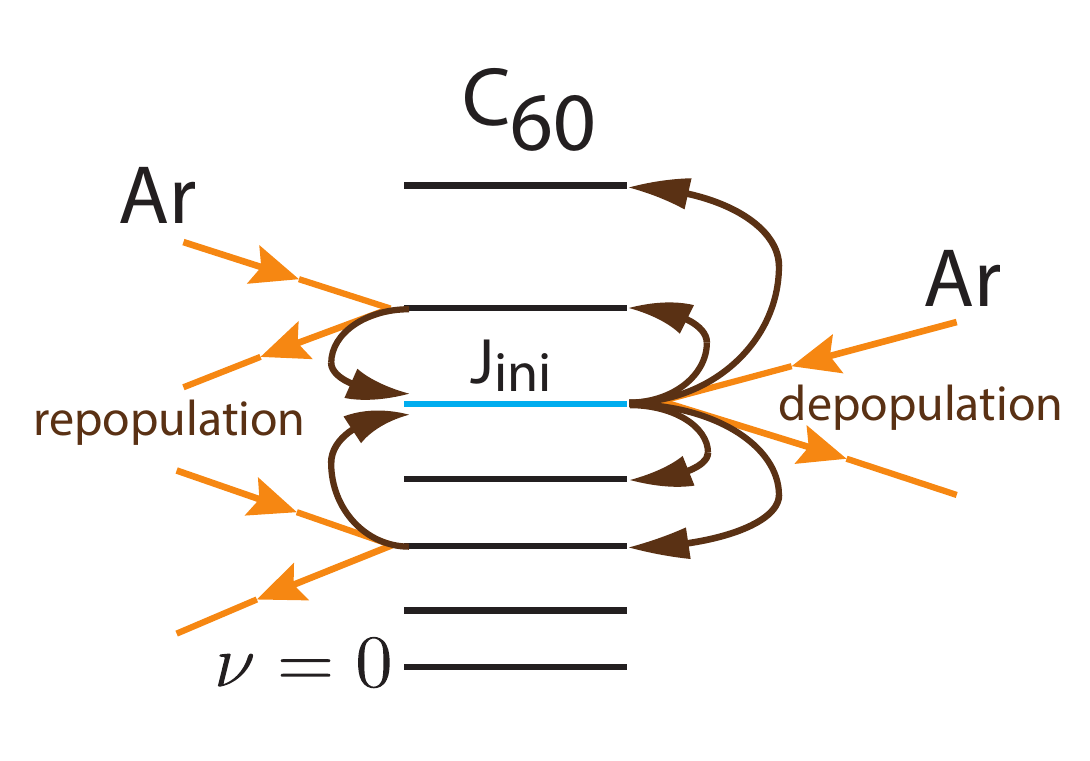} 
    \caption{Schematic of depopulation (quenching)  and repopulation  of  rotational states (black and blue horizontal lines) of the energetically lowest vibrational state of $^{12}$C$_{60}$ in collisions with argon atoms.}
\label{schematics}
\end{figure} 

The purpose of this article is to put forward a quantitative, quantum description of the  role of the molecular icosahedral $I_{\rm h}$ symmetry of $^{12}$C$_{60}$  in collisions with noble gas atoms and compute inelastic rate coefficients describing transitions among rotational states of $^{12}$C$_{60}$. Figure \ref{schematics} shows  the principle of these efforts. Optical pumping removes all population in a single rotational state of  $^{12}$C$_{60}$  in its ground vibrational level, denoted by angular momentum $J_{\rm ini}$, which is then  repopulated as well as  depopulated  in collisions with Ar rare-gas atoms. Because fullerenes have the highest possible point-group symmetry, this corresponds to partially uncharted territory in molecular collisions. We focus on $^{40}$Ar atoms as the collision partner and assume that  $^{12}$C$_{60}$ and the argon  atoms are cooled to temperatures between 100~K and 200~K. At these temperatures, we need only to concentrate on the thermal population of rotational levels of the $v=0$ vibrational state of $^{12}$C$_{60}$. The first excited vibrational mode of $^{12}$C$_{60}$ is
at energy $hc\times 273$ cm$^{-1}$ or $k\times 393$~K \cite{Vassallo1991}. Here, $h$ is the Planck constant, $c$ is the speed of light in vacuum, and $k$ is the Boltzmann constant.
We find that the total collisional cross-section and rate coefficient are  dominated by  elastic scattering from the isotropic short- and long-range interaction potential.

In Section~\ref{pes}, we describe our calculation of the C$_{60}$-Ar ground-state electronic potential energy 
surface (PES) and its expansion into anisotropic contributions.  
The allowed rotational states in gas phase $^{12}$C$_{60}$ are  affected by the   
icosahedral $I_{\rm h}$ symmetry and the fact that carbon-12 atoms are spin-zero bosons.  This is discussed  in Sec.~\ref{symmetry}.
In Sec.~\ref{sec:PT2}, we describe the required symmetry-adapted second-order perturbation theory to compute 
inelastic cross sections and rate coefficients at collision energy $E$.
We show in Sec.~\ref{rates} that inelastic, quenching rate coefficients of rotational levels  as high as $J=250$  are orders of magnitude smaller than elastic rate coefficients at the same $J$. Moreover, we predict collisional selection rules  that are different from those for molecules with lower symmetries. We summarize and conclude in Sec.~\ref{summary}.
In Appendix~\ref{polariz} we compute the dynamic dipole polarizability of  C$_{60}$  both as function of real and imaginary frequency. The behavior as function of imaginary frequencies is used in  calculations of dispersion coefficients between fullerenes and argon atoms.

\section{Isotropic and anisotropic components of the C$_{60}$-Ar potential energy surface} \label{pes}

We start this paper by describing our  {\it ab-initio} density functional theory (DFT) calculations using Gaussian 09 \cite{Gaussian09_RevE} employing the hybrid wB97XD functional where the long-range interactions are described using Grimme's D2 dispersion model parameter and the 6-31G(d,p) basis set. We use the counterpoise correction scheme to account for basis set superposition errors.
We have chosen this functional and dispersion model as a compromise between the quality of the potential energy surface and computational time to scan on a grid of spatial geometries. See Ref.~\cite{Tsuzuki2020} for an analysis of the accuracy of different dispersion models.

\begin{figure}
\includegraphics[scale=0.4,trim=0 0 0 0,clip]{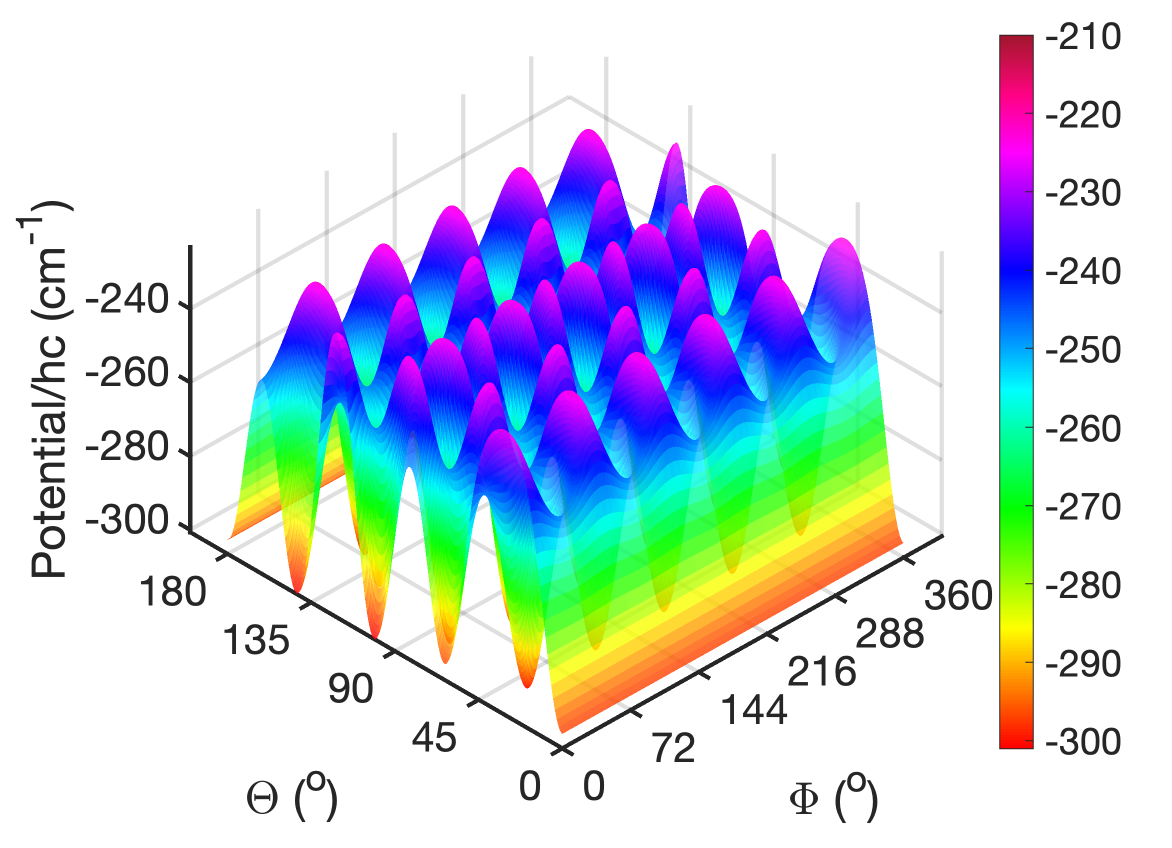}
\caption{Surface plot of the ground-state C$_{60}$-Ar potential as a function of  Jacobi angles $(\theta, \phi)$ for the equilibrium separation $R=13.6 a_0$ between the centers of masses of $^{12}$C$_{60}$ and $^{40}$Ar. }
\label{fig:C60Ar_3D_PES}
\end{figure}

We model the fullerene molecule  as a rigid body with the sixty carbon atoms fixed at their equilibrium positions, 
so that the interaction of the noble gas atom is described by  Jacobi coordinates $(R,\theta,\phi)$, 
where separation $R$ and body-fixed  angles  $\theta$ and  $\phi$ locate the center of mass of the $^{40}$Ar atom with respect to the center of mass of $^{12}$C$_{60}$. The polar $\theta$ and azimuthal $\phi$ angles are defined with respect to a right-handed Cartesian coordinate system with axes $x$, $y$, and $z$, where the $x$ and $z$ axes  coincides with a two- and five-fold symmetry axis  of $^{12}$C$_{60}$, respectively.  
Figure~\ref{fig:C60Ar_3D_PES} shows the two-dimensional cut through the PES at $R=13.6a_0$, where $a_0$ is the Bohr radius.
At this separation the potential has its absolute minima at $\approx hc\times -300$~cm$^{-1}$ relative to the potential energy for $R\to\infty$.  In fact, there are sixty equally deep minima. A depth of $hc \times 300$~cm$^{-1}$ is  a rather shallow bond and is on the order of our collision energies.
The anisotropic contributions, that is the $\theta$ and  $\phi$
dependence of the PES, are relatively small of order $ hc\times 20$~cm$^{-1}$. 
For separations $R<12.5a_0$, not shown in the figure, the PES is  repulsive. 

\begin{figure}
\includegraphics[scale=0.27,trim=0 0 0 0,clip]{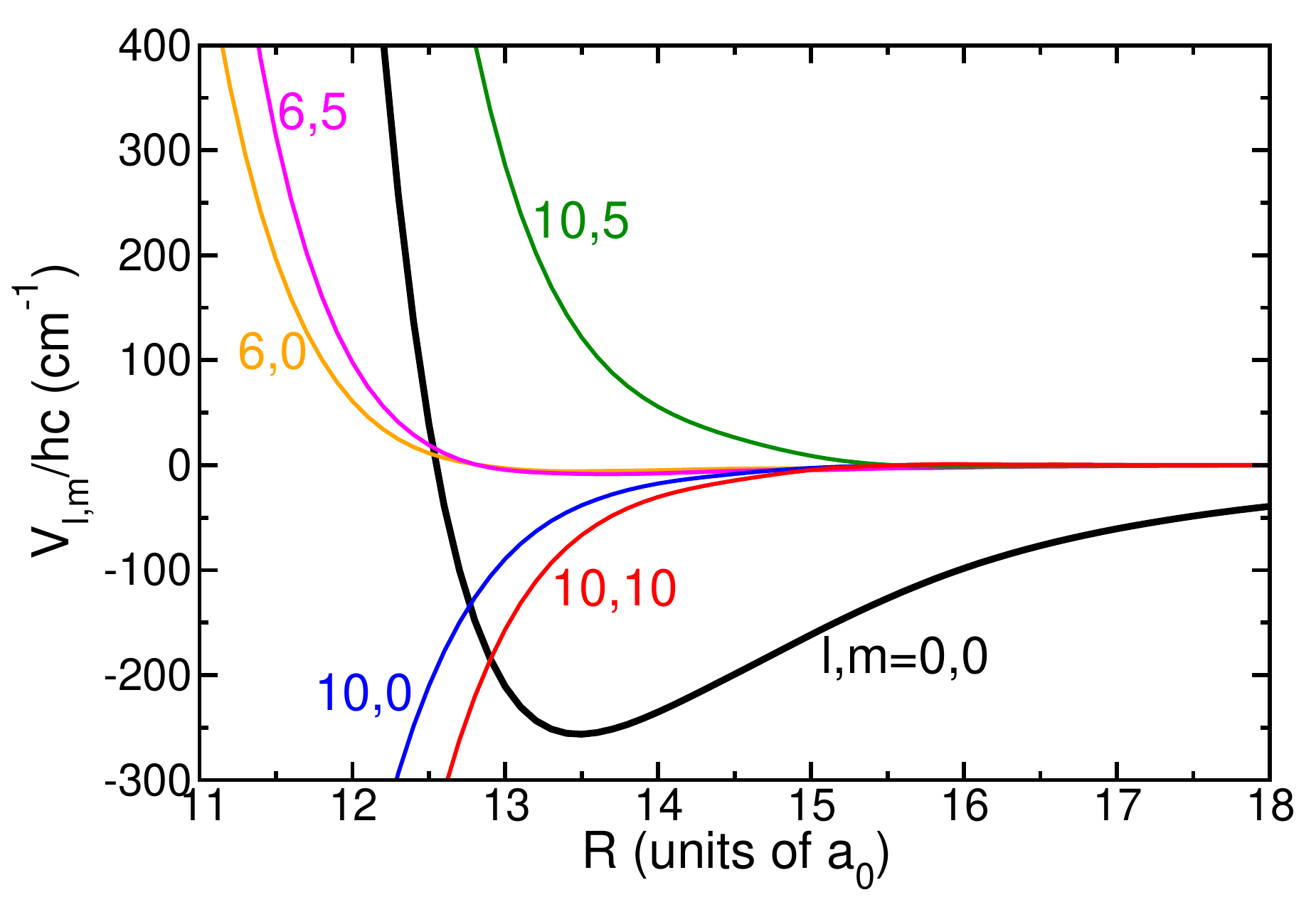}
\caption{Dominant non-zero radial expansion coefficients $V_{l,m}(R)$ of  C$_{60}$-Ar as functions of the separation  between the centers of masses of C$_{60}$ and Ar. Reproduced with permission from Ref.~\cite{Liu2022}.}
\label{coefficients}
\end{figure}

For  scattering  computations, we rewrite the PES for C$_{60}$-Ar   as a sum of Racah-normalized spherical harmonics $C_{lm}(\theta,\phi)$ that fulfill condition $C_{lm}(0,0)=\delta_{m0}$ and $R$-dependent radial strengths or coefficients $V_{l,m}(R)$. 
For the icosahedral symmetry group  this implies 
\begin{equation}
V(R,\theta,\phi)=\sum_{l,m\in{\cal L}} V_{l,m}(R)\frac{C_{lm}(\theta,\phi)+C_{l\, -m}(\theta,\phi)}{2}
\label{eq:expansion}
\end{equation}
in our body-fixed coordinate system,
where  quantum numbers $l, m$ are taken from the set $\cal L$ with $l=0,6,10,12,16,18, 20, ...$ and non-negative  $m=5n$ with $n=0,1,2,\cdots$ and $m\le l$ \cite{Harter1986, Harter1989, Saito1994, Bunker1999}. In practice, the coefficients are  obtained from a least squares fit to the {\em ab-initio} PES and we expand our potential with terms up to $l=20$ and  $m = 20$. The strongest expansion coefficients $V_{l,m}(R)$ as functions of $R$ are shown in Fig.~\ref{coefficients}.
The isotropic term $V_{0,0}(R)$ dominates in the attractive part of the PES.

Atom-molecule collisional dynamics is often controlled by its van der Waals interactions. We have used the Casimir-Polder formula \cite{CP} to determine the relevant $C_6$ van der Waals coefficients from  dynamic polarizabilities of C$_{60}$ and Ar as functions of imaginary frequencies.
Our computation of the dynamic polarizability of C$_{60}$ at its equilibrium geometry
as functions of both real and imaginary frequency are described in Appendix~\ref{polariz}, while data for Ar have been taken from Ref.~\cite{DEREVIANKO2010323}. 
We find that the isotropic $C_6$ coefficient is  $(2523\pm 250) E_{\rm h}a_0^6$, whereas the anisotropic dispersion coefficients are negligibly small. Here, $E_{\rm h}$ is the Hartree energy. Small  anisotropic  dispersion coefficients might suggest that  rotational quenching of $^{12}$C$_{60}$ in collisions with Ar will have a very low probability of occuring. As we will show, however, when our two particles approach each other, other anisotropies  induce small but non-negligible rotational transitions. 

\section{Icosahedral symmetry adapted rotational states of C$_{60}$} \label{symmetry}

The icosahedral symmetry  group $I_{\rm h}$ and Bose statistics among the sixty spinless $^{12}$C nuclei imply that only fully symmetric rotational wavefunctions $|JM,K\rangle$ 
are allowed
for the ground electronic-vibrational state with its zero total  electron angular momentum \cite{Harter1986, Harter1989, Saito1994, Bunker1999}.
These rotational states are eigenstates of the total angular momentum $\bf J$ operator of $^{12}$C$_{60}$ and  its projection operator $J_Z$ on the  $Z$ axis
in a space-fixed coordinate system $XYZ$ with corresponding quantum numbers $J$ and $M$, respectively. Fully symmetric implies that
\begin{equation}
\hat{\cal C}_i |JM,K\rangle = |JM,K\rangle\ {\rm and}\ 
 \hat{{\cal I}}\hat{\cal C}_i |JM,K\rangle = |JM,K\rangle,
\label{NWF}
\end{equation}
for all $i=1,\dots, 60$, where the $\hat{\cal C}_i$ are the sixty rotations around the center of mass of $^{12}$C$_{60}$ belonging to the rotational icosahedral  group $I$ and  ${\hat {\cal I}}$ is the spatial inversion operator. 
For given $J$ and $M$, index $K=1,\dots, N_{\rm h}(J)$  labels the  fully symmetric rotational wave functions.  The number of allowed states, 
$N_h(J)$, has been computed {\it a priori} using group theory (See for example Ref.~\cite{Harter1989}), is either 0 or 1 for  $J<30$, and  $N_{\rm h}(J)=N_{\rm h}({J-30})+1$ for $J\ge 30$.
In fact, we have $N_{\rm h}(J)= 1$,
       0,
       0,
       0,
       0,
       0,
       1,
       0,
       0,
       0,
       1,
       0,
       1,
       0,
  and 0 for $J=0,1,\dots,14$, respectively,
  and $N_{\rm h}(J)=1-N_{\rm h}(29-J)$ for $J<30$.
   The $N_{\rm h}(J)$ states correspond to the so-called superfine structure of $^{12}$C$_{60}$. 

The general form of the unit-normalized rotational eigenstates  is
\begin{equation}
|JM,K\rangle= \sum_{N=-J}^J a^{(JM)}_{KN}\,\sqrt{\frac{2J+1}{8\pi^2}} D^{J*}_{MN}(\phi_{\rm sf},\theta_{\rm sf},\varphi_{\rm sf}) 
\,,
\label{eq:general}
\end{equation}
where  Wigner rotation matrices $D^{J}_{MN}(\alpha,\beta,\gamma)$ are rigid
rotor solutions that are eigenstates of $J^2$, $J_Z$, and $J_z$  with quantum numbers $J$, $M$, and $N$, respectively \cite{BS1993}.
Euler angles  $\phi_{\rm sf},\theta_{\rm sf},\varphi_{\rm sf}$ orient the body-fixed $xyz$ axes of $^{12}$C$_{60}$, 
defined in Sec.~\ref{pes}, in the space-fixed $XYZ$ coordinate system. Superscript $*$ on the Wigner rotation matrices represents the complex conjugate.

The coefficients $a^{(JM)}_{KN}$ for each pair $J,M$ are uniquely determined by Eqs.~(\ref{NWF}). 
In fact, group theory states \cite{Harter1989} that the coefficients $a^{(JM)}_{KN}$ for  fully symmetric rotational states 
are given by the $N_h(J)$ linearly independent combinations of the $2J+1$ wavefunctions 
\begin{equation}
  \sum_{i=1}^{60} {\cal C}_i \,\sqrt{\frac{2J+1}{8\pi^2}} D^{J*}_{M,N}(\phi_{\rm sf},\theta_{\rm sf},\varphi_{\rm sf}) \,,
\label{FullS}
\end{equation}
one for each $N=-J,\dots,J$. 
The sum in Eq.~(\ref{FullS}) is over all sixty rotation operators ${\cal C}_i$ in the  rotational icosahedral group and gives each Wigner rotation matrix  equal weight.
(These weights are the characters of the fully symmetric irreducible representation and we recall that $N_{\rm h}(J)\le 2J+1$.)
The action of a rotation $\hat{\cal C}_i$  on Wigner functions is given by
\begin{eqnarray}
\lefteqn{\hat{\cal C}_i  D^{J*}_{M,N}(\phi_{\rm sf},\theta_{\rm sf},\varphi_{\rm sf}) =} \label{eq:action}\\
&& \quad\sum_{N'=-J}^{J} D^{J*}_{M,N'}(\phi_{\rm sf},\theta_{\rm sf},\varphi_{\rm sf})D^{J}_{N',N}(-\gamma_i,-\beta_i,-\alpha_i) \,,
\nonumber
\end{eqnarray}
 where $\alpha_i,\beta_i,\gamma_i$ are the Euler angles describing the $i$-th rotation. 

 Finally, the inversion operator  ${\cal I}$ leaves Wigner rotation matrices unchanged up to phase factor $(-1)^J$ and, 
thus,   multiplies each term in Eq.~(\ref{FullS}) by a factor of two. We omitted this factor in the equation as it only changes the normalization.
In practice, for each $J,M$ we first construct all $2J+1$ states in Eq.~(\ref{FullS}) and then use  singular value decomposition
to determine all linearly independent (and orthonormal) states. Here, we rely on the list of Euler angles for the 60 rotations from Ref.~\cite{Martinez1989}, a rotation between our coordinate system and that of Ref.~\cite{Martinez1989},
and the numerical evaluation of Wigner rotation matrices, {\it e.g.}, see Refs.~\cite{Dachsel2006,Tajima2015,Wang2022}. This enables us to independently verify the correct  $N_{\rm h}(J)$.
When $N_{\rm h}(J)=0$ for some of the $J<30$, Eq.~(\ref{FullS}) is  zero for all $N$.

We start our analysis of collision-induced excitation and quenching of rotational states of $^{12}$C$_{60}$ in its ground vibrational state by 
studying the population distribution of  rotational levels $J$  of $^{12}$C$_{60}$ for the three temperatures $T=  
100$~K, 150~K, and 300~K. The distributions are given by 
\begin{equation}
p (J;T)=(2J+1)N_{\rm h}(J)e^{-BJ(J+1)/kT}/Z(T)\,,
\label{eq:Rot_Dist}
\end{equation}
where $B=hc\times0.0028$ cm$^{-1}$ is the rotational constant of $^{12}$C$_{60}$ in its ground vibrational state,  obtained from a fit to the experimental spectra in Refs.~\cite{Changala2019,Lee:2023},  $Z(T)$ is a normalization factor such that $\sum_{J=0}^\infty p(J;T)=1$.
For our study of collisional physics, we do not need to account for  small energy shifts due to centrifugal distortions and the superfine structure of $^{12}$C$_{60}$.

\begin{figure}
\includegraphics[scale=0.24,trim=0 0 0 0,clip]{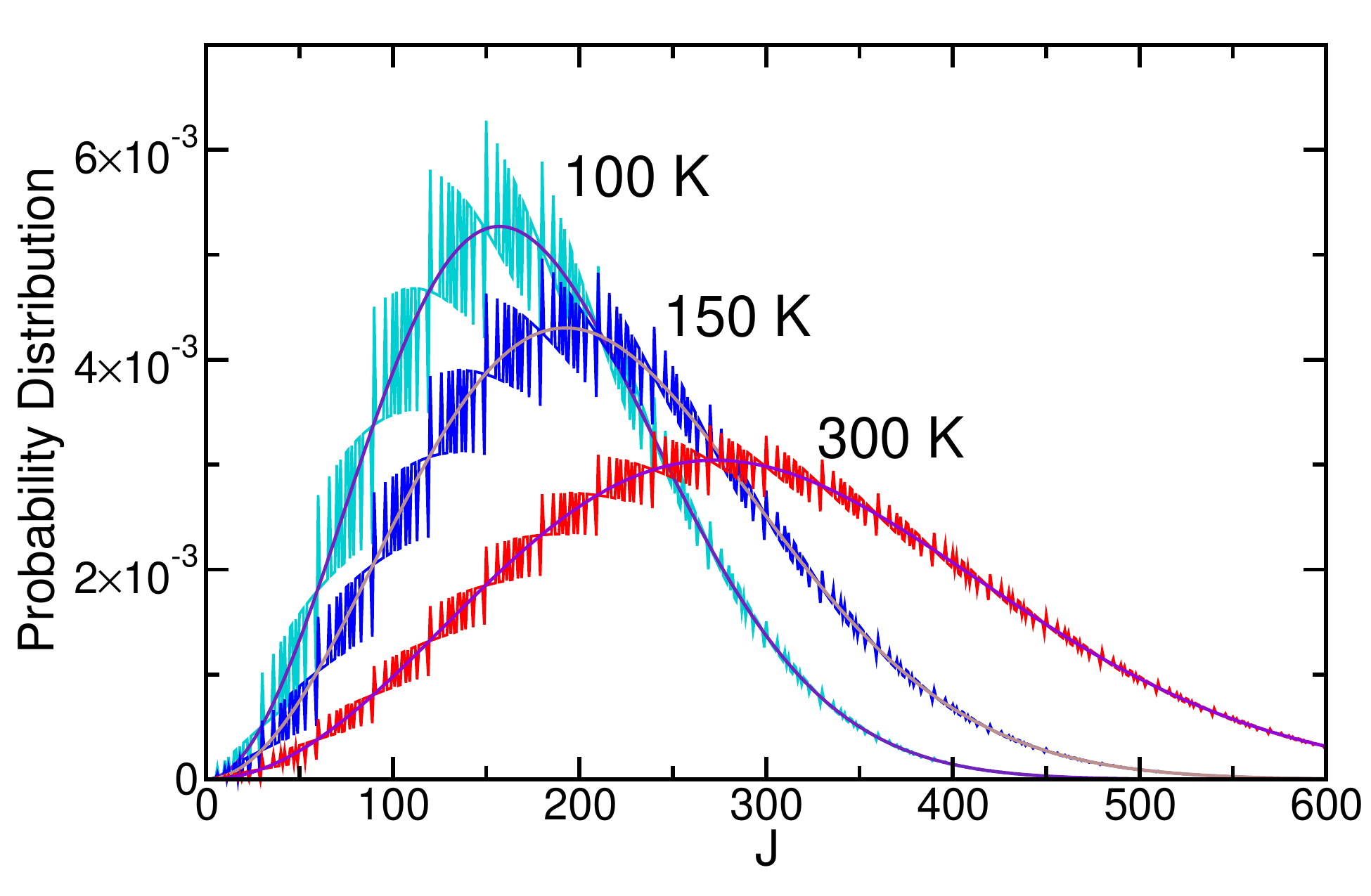}
    \caption{Population or probability distributions of rotational levels $J$ of the ground vibrational state of the $^{12}$C$_{60}$ fullerene at temperatures of $T=100$~K (cyan curve), 150~K (blue curve) and 300~K (red curve). 
The smooth solid lines going through the center of jagged curves are rotational distributions that treat the $^{12}$C$_{60}$ as a rigid spherical top, for which the degeneracy factor is $(2J+1)^2$ for each $J$. 
}
\label{rot_distribution}
\end{figure}

Figure~\ref{rot_distribution} shows distributions as functions of $J$ for the three temperatures.
We observe that for $T=150$~K angular momenta $J$  up to $\approx350$ have significant occupation.
In addition, for each temperature the figure shows the cases where the  icosahedral symmetry of $^{12}$C$_{60}$ is accounted for and  one where the fullerene is modeled as a spherical top molecule, {\it i.e.}, where $N_{\rm h}(J)$ is replaced by $2J+1$.  The jagged distributions, when the  icosahedral symmetry are accounted for, are a consequence of the complex  degeneracy factor $N_{\rm h}(J)$ for the number of the allowed states at a given $J$. 
The jaggedness becomes less pronounced both with increasing quantum number $J$  and increasing temperature and the curves approach the distribution of a rigid spherical top molecule. 
 
\section{Theory for cross sections for rotational quenching of C$_{60}$ by Ar atoms}\label{sec:PT2}

We use second-order perturbation theory (PT2) to compute  inelastic cross sections for rotational transitions $J,K \rightarrow J',K'$  averaged over both the direction $\hat k$ of wave\-vector {\bf k} in the relative motion of the $^{12}$C$_{60}$-$^{40}$Ar system with reduced mass $\mu$  and the initial projection $M$ of the angular momentum $J$ of $^{12}$C$_{60}$ in its vibrational and electronic ground state.
We account for both positive and negative values of  quantum number $K$
and, thereby,  have not taken advantage of the parity symmetry inherent in the system. 
The zeroth-order Hamiltonian is defined by the relative motion of $^{12}$C$_{60}$-$^{40}$Ar in the isotropic potential $V_{0,0}(R)$.
This zeroth-order Hamiltonian conserves the relative orbital angular momentum or partial wave ${\bm\ell}$ between $^{12}$C$_{60}$ and $^{40}$Ar.
Within PT2, inelastic cross sections are then given by
\begin{eqnarray}
\lefteqn{\sigma_{J'K',JK}(E) = \frac{1}{4\pi} \int{d\hat{k}}\frac{1}{2J+1} \sum_{MM'} \frac{2\pi}{\hbar}}\label{sigma1}\\
 &&\times\sum_{\ell'=0}^{\infty}  \sum_{m'=-\ell'}^{\ell'} \left|\left<\psi^{\bf k}_{JM,K}\left| V(R,\theta,\phi) \right| \psi^E_{\ell'm',J'M'K'} \right> \right|^2,
\nonumber
\end{eqnarray}
where 
\begin{equation}
  |\psi^{\bf k}_{JM,K}\rangle = 4\pi \sqrt{\frac{\mu}{\hbar k}} |JM,K\rangle \sum_{\ell=0}^{\infty}i^\ell f_{\ell k}(R) 
  \left(Y_{\ell}(\hat{k})\cdot Y_{\ell}(\hat{R})\right)
\end{equation}
are unit flux normalized, zeroth-order radial wavefunctions
and 
\begin{eqnarray}
 |\psi^E_{\ell m,JMK}\rangle = |JM,K\rangle \sqrt{\frac{2k\mu}{\pi \hbar^2}} f_{\ell k}(R) Y_{\ell m}(\hat{R})
\end{eqnarray}
are energy normalized, zeroth-order radial wavefunctions.
Furthermore, kets $|JM,K\rangle$ are defined by Eq.~(\ref{eq:general}), $Y_{\ell m}(\hat{x})$ are unit-normalized spherical harmonic functions, and $f_{\ell k}(R)$ are solutions of the radial Schr\"odinger equation
\begin{equation}
\left( -\frac{\hbar^2}{2\mu}\frac{d^2}{dR^2} + \frac{\hbar^2\ell(\ell+1)}{2\mu R^2}+ V_{0,0}(R) \right) f_{\ell k}(R) = E f_{\ell k}(R),
\label{flkeq}
\end{equation}
with kinetic energy $E=\hbar^2k^2/(2\mu)$, partial wave $\ell $, and asymptotic behavior 
\begin{equation}
 f_{\ell k}(R) \rightarrow \frac{1}{k}\sin\left(kR-\pi \ell /2+\eta_\ell (k)\right)
\label{asymp}
\end{equation}
for $R\to\infty$. Here, $\eta_\ell(k)$ are  phase shifts.

After integration over $\hat{k}$ Eq.~(\ref{sigma1}) becomes
\begin{eqnarray}
\nonumber
\sigma_{J'K',JK}(E) =  \frac{1}{2\hbar} \frac{1}{2J+1} \sum_{MM'} \sum_{\ell,\ell'=0}^{\infty}  \sum_{m=-\ell}^{\ell} \sum_{m'=-\ell'}^{\ell'} \\
  \left|\left<\psi^{(1)}_{\ell m,JMK}\left| V(R,\theta,\phi) \right| \psi^E_{\ell'm',J'M'K'} \right> \right|^2,
\label{sigma2}
\end{eqnarray}
where
\begin{eqnarray}
  \psi^{(1)}_{\ell m,JMK} = 4\pi \sqrt{\frac{\mu}{\hbar k}} |JM,K\rangle i^\ell f_{\ell k}(R) Y_{\ell m}(\hat{R}).
\end{eqnarray}
Note that $m+M=m'+M'$ is conserved in the collision.

Rather than numerically solving Eq.~(\ref{flkeq}) by a finite difference method,  we  solve this equation  using a discrete variable representation (DVR) \cite{Colbert1992} to discretize  the differential operator $d^2/dR^2$
on radial interval $[R_{\rm min},R_{\rm max}]$ with $R_{\rm min}=11a_0$, $R_{\rm max}=1000a_0$, and discretization step $\delta R=0.1a_0$.
In the DVR representation, $f_{\ell k}^{\rm DVR}(R_{\rm min})=f_{\ell k}^{\rm DVR}(R_{\rm max})=0$ and we find a set of discrete energies $E$ that
satisfy these conditions by solving the relevant eigenvalue problem  for each $\ell$.
The unit-normalized eigensolutions $f_{\ell k}^{\rm DVR}(R)$ are converted to
wavefunctions with normalization given in Eq.~(\ref{asymp}) using
\begin{equation}
f_{\ell k}(R)=  \frac{1}{k}\sqrt{\frac{R_{\rm max}}{2}} f_{\ell k}^{\rm DVR}(R) .
\end{equation}
Matrix elements in Eq.~(\ref{sigma2}) at arbitrary $E$ can be found by interpolation.
We have included several hundreds of partial waves $\ell$ to reach convergence within our perturbation theory.

\section{Collision-induced quenching and repopulation}\label{rates}

\begin{figure*}
\includegraphics[scale=0.30]{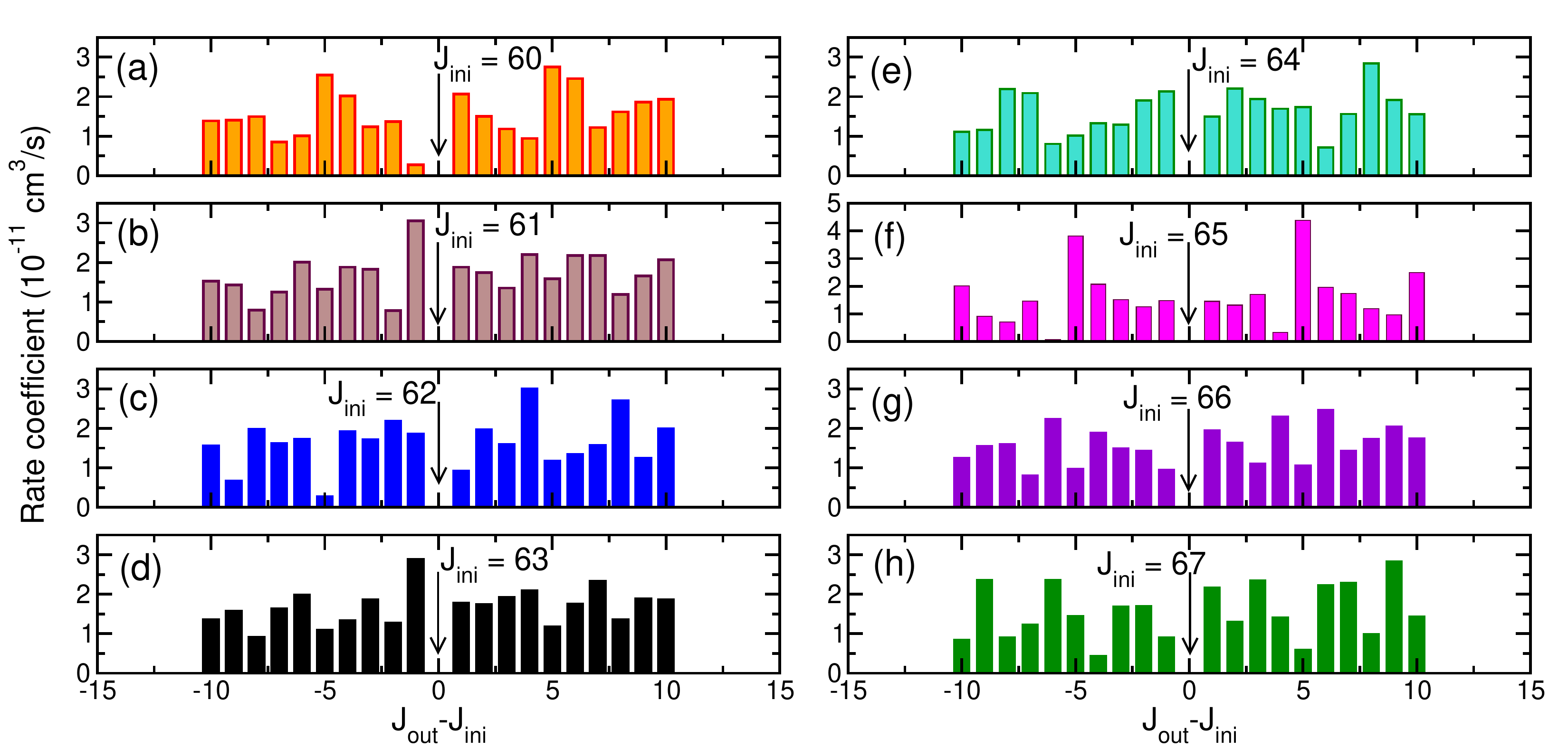}
\caption{Collision-induced quenching and repopulation rate coefficients of rotational levels $J_{\rm ini}$ of $^{12}$C$_{60}$ between 60 and 67 (panels (a) through (h)) with $^{40}$Ar as  functions of the final rotational state $J_{\rm out}$ at collisional energy $E = k\times 150$~K. The vertical axes are the same for all panels except that for panel (f). }
\label{fig:j60to70}
\end{figure*}

In this section, we present perturbative results for rotational inelastic rate coefficient 
\begin{equation}
K_{J',J}(E)=\frac{1}{N_{\rm h}(J)}\sum_{K=1}^{N_{\rm h}(J)} \sum_{K'=1}^{N_{\rm h}(J')}\sum v\sigma_{J'K',JK}(E)\,,
\end{equation}
where relative velocity $v=\hbar k/\mu$. Figure \ref{fig:j60to70}  shows perturbative inelastic rate coefficients
at collision energy $E=k\times 150$~K
for sequential initial angular momenta $J_{\rm ini}$ of $^{12}$C$_{60}$ between 60 and 67
as functions of outgoing and final angular momenta $J_{\rm out}\ne J_{\rm ini}$. We have only included the contributions
from the strongest expansion coefficients $V_{l,m}(R)$ of the  interaction potential $V(R,\theta,\varphi)$ shown in Fig.~\ref{coefficients}. Hence, the outgoing  angular momentum satisfies $J_{\rm ini}-10 \le J_{\rm out}\le J_{\rm ini}+10$.
We also observe that rate coefficients 
have a non-monotonic, seemingly random behavior as functions of $J_{\rm out}$ around a  value just under  $2\times 10^{-11}$ cm$^3$/s both for $J_{\rm out}<J_{\rm ini}$ corresponding to rotational de-excitation and for $J_{\rm out}>J_{\rm ini}$ corresponding to rotational excitation. In fact, no obvious pattern from one initial angular momentum to another exists,
The complex behavior is due to the $I_{\rm h}$ symmetry of $^{12}$C$_{60}$.

\begin{figure}
\includegraphics[scale=0.30]{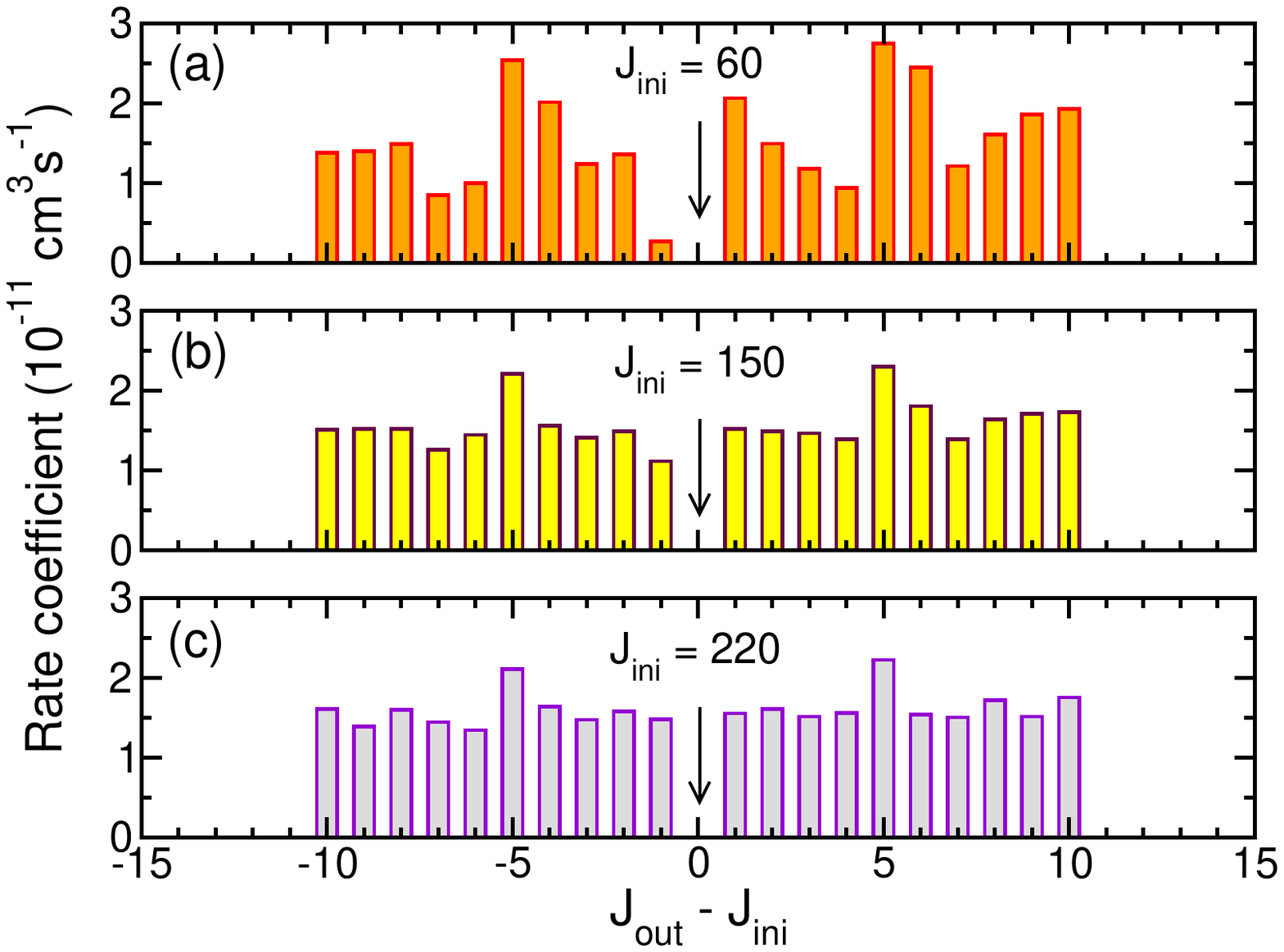}
\caption{Collision-induced inelastic  rate coefficients of initial rotational levels $J_{\rm ini}=60$, 150, and 220 of $^{12}$C$_{60}$ (panels (a), (b), and (c), respectively) with $^{40}$Ar  as  functions of the final rotational state $J_{\rm out}$ at collision energy $k\times 150$~K. }
\label{fig:j60to220}
\end{figure}

Figure \ref{fig:j60to220} shows perturbative rate coefficients at collision energy $E=k\times 150$~K for initial angular momenta $J_{\rm ini}=60$, 150, and 220 spanning the relevant significantly populated angular momenta at temperature $T=150$~K. Several observations can be made. Firstly,
the amplitude of the ``random'' behavior of the rate coefficients is smaller for larger  $J_{\rm ini}$. Secondly, the pattern as function
of $J_{\rm out}-J_{\rm ini}$ is very similar for the three $J_{\rm ini}$ with the largest rate coefficients occurring for 
$|J_{\rm out}-J_{\rm ini}|=5$.

The elastic rate coefficients with $J_{\rm out}=J_{\rm ini}$ at the same collision energy (not shown) are more than a hundred times larger and within our approximate approach derived for scattering from the isotropic potential $V_{0,0}(R)$ and, thus, phase shifts $\eta_\ell(k)$.  In
fact, this elastic rate coefficient is very close to its semi-classical value from Ref.~\cite{child:1974} given by
\begin{equation}
      K_{\rm elast}(E)=  6.125\cdots  \left(\frac{E}{E_6}\right)^{3/10} \frac{\hbar x_6}{\mu}\,,
\end{equation}
where van der Waals energy $E_6=\hbar^2/(2\mu x_6^2)$   and van der Waals  length $x_6=\sqrt[4]{2\mu C_6/\hbar^2}$.
See, for example, Ref.~\cite{Klos2023} for a comparison of thermalized total elastic rate coefficients using quantum simulations and the semi-classical model.
For our system, $E_6\approx k\times 0.12$~mK and $\hbar x_6/\mu\approx 1.2\times 10^{-11}$~cm$^3$/s,
so that $K_{\rm elast}(E)\approx 5.0\times 10^{-9}$ cm$^3$/s at $E/k=150$~K.
In other words, inelastic rate coefficients are small, validating our perturbative procedure for these rate coefficients.

\begin{figure}
\includegraphics[scale=0.30]{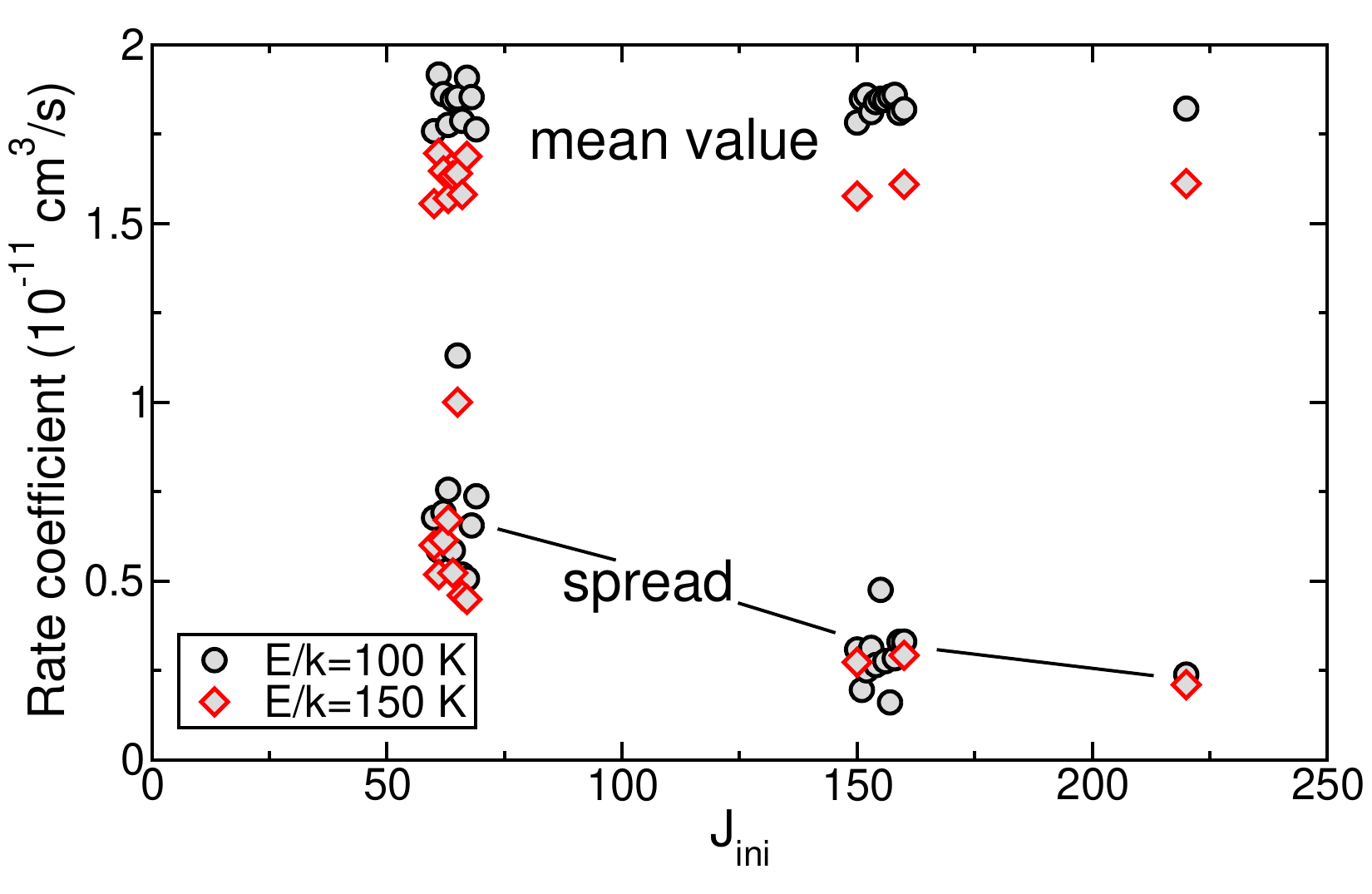}
\caption{The mean and spread of $^{12}$C$_{60}$-$^{40}$Ar collision-induced inelastic rate coefficients as function of initial rotational level $J_{\rm ini}$ of $^{12}$C$_{60}$ at collision energies $k\times 100$~K (black circles) and $k\times 150$~K (red diamonds). The  mean and spread have been computed with respect to final rotational state $J_{\rm out}$.}
\label{fig:mean}
\end{figure}

Finally, Fig.~\ref{fig:mean} shows the mean and spread of collision-induced inelastic rate coefficients between $^{12}$C$_{60}$ and $^{40}$Ar 
as a function of  initial angular momentum $J_{\rm ini}$ of $^{12}$C$_{60}$ at  collision energies $k\times 100$~K and $k\times 150$~K. Within our perturbative calculations, these quantities are defined as
\begin{equation}
 {\cal K}_{J}(E)= \frac{1}{20} \sum_{J'=J-10, J'\ne J}^{J+10} K_{J',J}(E)
 \end{equation}
and
\begin{equation}
 u_{J}(E)=\sqrt{ \frac{1}{20} \sum_{J'=J-10, J'\ne J}^{J+10} \left(K_{J',J}(E)-{\cal K}_{J}(E)\right)^2}\,,
\end{equation}
respectively. We note that ${\cal K}_{J_{\rm ini}}(E)$ and $u_{J_{\rm ini}}(E)$ have only been computed for a limited set of $J_{\rm ini}$ due to the
combined complexity of computing symmetry-adapted states $|JM,K\rangle$, eigensolutions $f_{\ell k}(R)$, matrix elements of potentials $V_{l,m}(R)$, 
and summing over initial and final partial wave $\ell$ and $\ell'$ and all projection quantum numbers.
 
We also observe that  mean  ${\cal K}_{J_{\rm ini}}(E)$ has small, seemingly random variations from one $J_{\rm ini}$ to the next that become
smaller for increasing $J_{\rm ini}$ as best observed for the data at $E=k\times 100$ K. Still the mean is mostly 
independent of $J_{\rm ini}$ over a larger range of $J_{\rm ini}$, {\it i.e.}, those relevant for our  temperatures. 
Finally, the mean value slowly  decreases with $E$ for energies around $k\times 100$~K and  $k\times150$~K. The spread $u_{J_{\rm ini}}(E)$ also has small, random variations from one $J_{\rm ini}$ to the next, but is observably getting smaller over larger ranges of $J_{\rm ini}$. 
The spread $u_{J_{\rm ini}}(E)$ is larger than the variation in mean  ${\cal K}_{J_{\rm ini}}(E)$ from one $J_{\rm ini}$ to the next.

Our observations regarding the weak dependence of ${\cal K}_{J}(E)$ and $u_{J}(E)$ on energy $E$ and $^{12}$C$_{\rm 60}$ rotational state $J$ in Fig.~\ref{fig:mean} enable an estimate
of thermalized rate coefficients for each $J$ and temperatures between 100~K and 150~K. We assume $\langle {\cal K}_{J}(T)\rangle\approx {\cal K}_J(E=kT)$ and we find typical thermalized inelastic rate coefficients of just under $2\times 10^{-11}$ cm$^3$/s.

For typical datasets used in Ref.~\cite{Liu2022}, the Ar number density was $n_{\rm Ar}=1.6\times 10^{16}$~cm$^{-3}$, which leads to 
a  time scale $\tau_{\rm inelas}\approx 3$~$\mu$s for removing population from total angular momentum $J_{\rm ini}$ of $^{12}$C$_{\rm 60}$
due to collisions with argon atoms. 
This time scale, however, cannot be directly compared to time scales measured in experiments  by Ref.~\cite{Liu2022} implementing the setup in Fig.~\ref{schematics}. Such a model must account for  laser beams with millimeter beam waists,  C$_{\rm 60}$ molecules diffusing in and out of these beams,
and the mean free path for elastic C$_{60}$-Ar collisions given by 
$\langle v\rangle/[\langle K_{\rm elast}(E)\rangle n_{\rm Ar}]\approx4$~$\mu$m at $T=150$~K. 
Here, the thermalized relative velocity $\langle v\rangle$ is given by $\langle v\rangle= (2/\sqrt{\pi}) \sqrt{2kT/\mu}$.
We might expect that the observed time scale will be many times $\tau_{\rm inelas}$. This modeling falls outside the scope of this paper.

\section{Summary}\label{summary}

We have described rotationally inelastic  collisions of the $^{12}$C$_{60}$ molecule with inert $^{40}$Ar  for gas temperatures near 150 K. At these 
temperature, we can concentrate on the thermal population of hundreds of rotational levels of the $v = 0$ vibrational state of $^{12}$C$_{60}$ in its 
electronic ground state. We summarized a  calculation of the anisotropic terms in the interaction potential energy surface of the system, described 
internal rotational states of  C$_{60}$ in terms of its  icosahedral symmetry, and performed second-order perturbative quantum calculations
of the rotationally inelastic inelastic rate coefficients. 
The  inelastic rate coefficients are found to be order of magnitude smaller than the elastic rate coefficients.  

\section*{Acknowledgements}
SK, AL,  and JK acknowledges funding by the AFOSR, Grant No. FA9550-21-1-0153, National Science Foundation, Grant No. PHY-2409425, and the Gordon and Betty Moore Foundation, Grant No. GBMF12330.

\appendix
\section{Dipole polarizability of C$_{60}$  and van der Waals coefficient for C$_{60}$-Ar} \label{polariz}

The C$_{60}$ molecule in its electronic and vibrational ground state has a high symmetry  and many of the lowest multipole moments vanish. For example,  dipole or quadrupole moments are zero. The dipole polarizability tensor $\alpha_{ab}(\omega)$ with ${a,b=x}$, $y$, and $z$ in the body-fixed coordinate system and laser frequency $\omega$ or a photon energy $\hbar\omega$, however, is non-zero. The isotropic dipole polarizability is defined by $\alpha_{\rm iso}(\omega)\equiv\sum_{a}\alpha_{aa}(\omega)/3$.

The isotropic, static dipole polarizability $\alpha_{\rm iso}(0)$ of the fullerene in its electronic and vibrational ground state has  been extensively  studied both experimentally and theoretically.
A recent experimental value  is $4\pi\epsilon_0\times 590(17) a_0^3$ \cite{Fein2019}, where the number in parentheses is the standard uncertainty  in the last two digits and $\epsilon_0$ is the vacuum electric permittivity. A recent theoretical value of $4\pi\epsilon_0\times 545.3 a_0^3$  including a small contribution from zero-point vibrational motion was published in Ref.~\cite{Lao2021}. 
Significant differences in values for $\alpha_{\rm iso}(0)$ with a spread larger than the experimental uncertainty of Ref.~\cite{Fein2019} among the published experimental and theoretical  evaluations, however, exists.

In Ref.~\cite{Liu2022} we  calculated the static dipole polarizability tensor of C$_{60}$ using density functional theory in the Q-Chem program with the 6-311G(d,p) basis and Perdew-Becke functionals \cite{Perdew1986, Becke1988}.  The calculations  lead to $\alpha_{\rm iso}(0)=4\pi\epsilon_0\times 502.6420a_0^3$, a value that is at the lower end of the published theoretical values. In addition, we derived a small anisotropy of the static polarizability tensor of $\kappa(0)=3.5\times10^{-6}$  defined via
\begin{equation}
\kappa^2(\omega)=
\frac{1}{6}\sum_a \left(\frac{\alpha_{aa}(\omega)-\alpha_{\rm iso}(\omega)}{\alpha_{\rm iso}(\omega)}\right)^2
\end{equation} 
for any frequency $\omega$.

\begin{figure}
\includegraphics[scale=0.31]{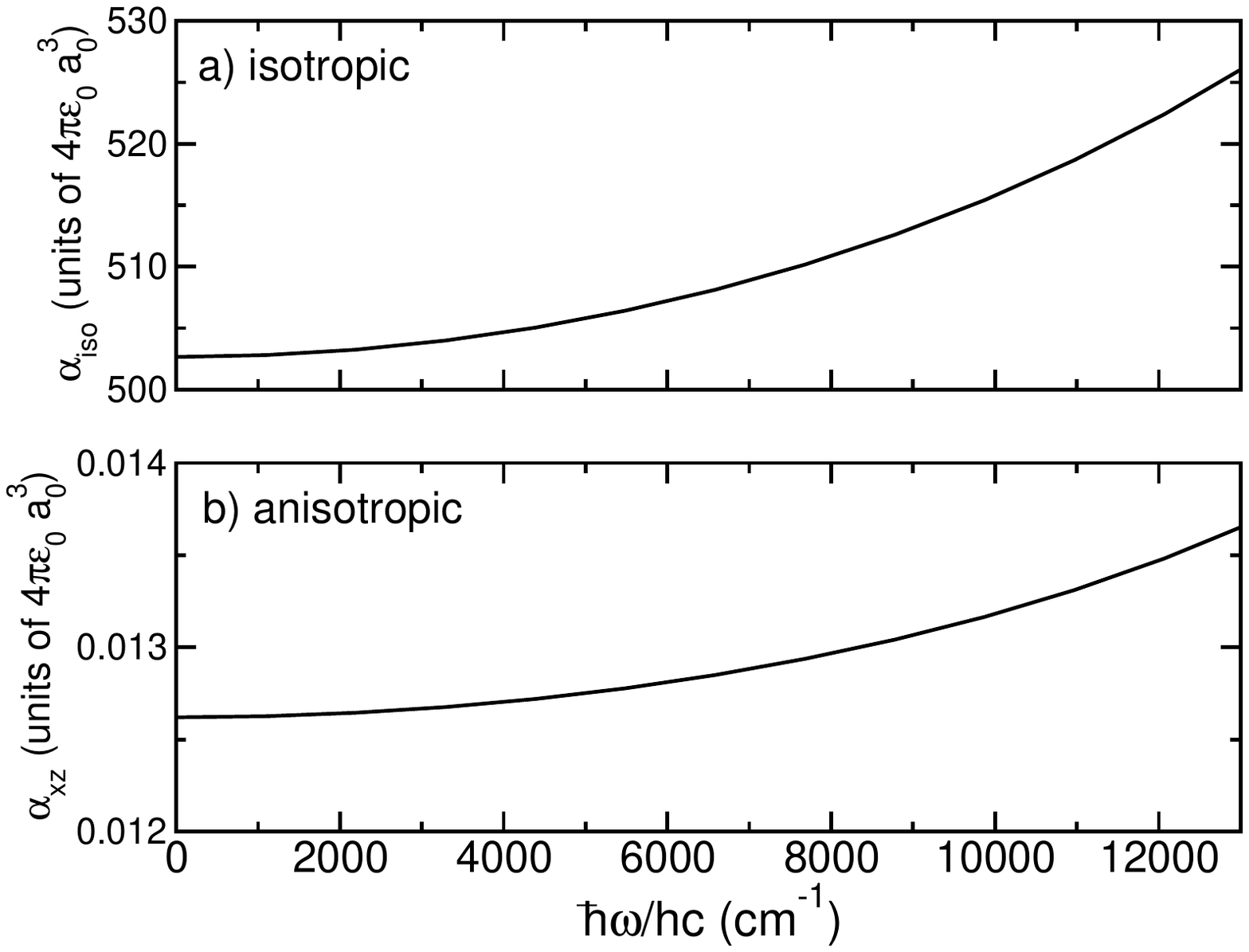}
\caption{Scaled isotropic $\alpha_{\rm iso}(\omega)$ (panel a) and anisotropic $\alpha_{xz}(\omega)$  (panel b) dynamic polarizability of C$_{60}$ as functions of laser frequency $\omega$.}
\label{fig:polarizability}
\end{figure}

In order to estimate the $C_6$ dispersion  coefficient for the C$_{60}$-Ar system, we also calculated the frequency-dependent dipole polarizability tensor using the coupled-cluster propagator implemented in the MOLPRO program \cite{molpro} with single and double excitations (CCSD) and the minimal STO-3G basis due to computational limitations
These calculations were performed up to $\hbar\omega=hc\times 13\,000$~cm$^{-1}$ beyond which the coupled-cluster propagator did not converge.
For each tensor component, the frequency-dependent results  for the dipole polarizability are then  scaled to coincide with the corresponding static dipole polarizability tensor components obtained by our Q-Chem calculations with its larger basis set. These scaled data are shown in Fig.~\ref{fig:polarizability} as function of frequency $\omega$. The dynamic dipole polarizability has a singularity at a photon energy $\Delta>hc\times 13\,000$~cm$^{-1}$ corresponding to a transition to an electronically excited state of C$_{60}$.

We fit the dynamic dipole polarizability to the functional form $\alpha_{ab}(\omega)=\alpha_{ab}(0)\Delta^2/[\Delta^2-(\hbar\omega)^2]$
with adjustable parameters $\alpha_{ab}(0)$ and $\Delta$
and use $\alpha_{ab}(i\omega)$ at imaginary frequencies $i\omega$
with the dipole polarizability of the Ar atom at imaginary frequencies  from Ref.~\cite{DEREVIANKO2010323} in the Casimir-Polder formula \cite{Stone2013}, to find that the isotropic $C_{6}$ dispersion coefficient is $2523E_{\rm h}a_0^6$. We estimate that this dispersion coefficient
has a standard uncertainty of 10\,\%.
The anisotropic dispersion coefficients are effectively zero.

\bibliography{c60arhe_references_2nd_paper.bib}

\end{document}